\documentclass[prd,twocolumn,letterpaper,superscriptaddress,nofootinbib]{revtex4}
\usepackage{amsmath,epsfig,natbib,bm,psfrag}

\begin{document}
\newcommand{\numfrac}[2]{{\textstyle \frac{#1}{#2}}}
\newcommand\ba{\begin{eqnarray}}
\newcommand\ea{\end{eqnarray}}
\newcommand\be{\begin{equation}}
\newcommand\ee{\end{equation}}
\newcommand\lagrange{{\cal L}}
\newcommand\cll{{\cal L}}
\newcommand\clx{{\cal X}}
\newcommand\clz{{\cal Z}}
\newcommand\clv{{\cal V}}
\newcommand\clo{{\cal O}}
\newcommand\cla{{\cal A}}
\newcommand{\grad}{\nabla}

\newcommand{\Psil}{\Psi_l}

\newcommand{\bsigma}{{\bar{\sigma}}}
\newcommand{\bI}{\bar{I}}
\newcommand{\bq}{\bar{q}}
\newcommand{\bv}{\bar{v}}

\newcommand\del{\nabla}
\newcommand\Tr{{\rm Tr}}
\newcommand\half{{\frac{1}{2}}}
\newcommand\fourth{{1\over 8}}
\newcommand\bibi{\bibitem}

\newcommand\calS{{\cal S}}
\renewcommand\H{{\cal H}}
\newcommand\K{{\cal K}}
\newcommand\opacity{\tau_c^{-1}}

\renewcommand\P{{\cal P}}

\newcommand{\la}{\langle}
\newcommand{\ra}{\rangle}

\newcommand{\Omtot}{\Omega_{\mathrm{tot}}}
\newcommand\xx{\mbox{\boldmath $x$}}
\newcommand{\phpr} {\phi'}
\newcommand{\gam}{\gamma_{ij}}
\newcommand{\sqgam}{\sqrt{\gamma}}
\newcommand{\delk}{\Delta+3{\K}}
\newcommand{\dph}{\delta\phi}
\newcommand{\om} {\Omega}
\newcommand{\dom}{\delta^{(3)}\left(\Omega\right)}
\newcommand{\rar}{\rightarrow}
\newcommand{\Rar}{\Rightarrow}
\newcommand{\labeq}[1] {\label{eq:#1}}
\newcommand{\eqn}[1] {(\ref{eq:#1})}
\newcommand{\labfig}[1] {\label{fig:#1}}
\newcommand{\fig}[1] {\ref{fig:#1}}
\newcommand\gsim{ \lower .75ex \hbox{$\sim$} \llap{\raise .27ex \hbox{$>$}} }
\newcommand\lsim{ \lower .75ex \hbox{$\sim$} \llap{\raise .27ex \hbox{$<$}} }
\newcommand\bigdot[1] {\stackrel{\mbox{{\huge .}}}{#1}}
\newcommand\bigddot[1] {\stackrel{\mbox{{\huge ..}}}{#1}}
\newcommand{\Mpc}{\text{Mpc}}

\newcommand{\curl}{\,\mbox{curl}\,}
\newcommand{\ord}{\mbox{O}}
\newcommand{\sigt}{\sigma_{\mathrm{T}}}
\newcommand{\nelec}{n_{\mathrm{e}}}
\newcommand{\ud}{{\mathrm{d}}}
\newcommand{\uD}{{\mathrm{D}}}
\newcommand{\tc}{t_{\mathrm{c}}}
\newcommand{\clq}{{\mathcal{Q}}}
\newcommand{\TT}{{\mathrm{TT}}}
\newcommand{\clh}{{\mathcal{H}}}
\newcommand{\clp}{{\mathcal{P}}}

\newcommand{\short}[2]{#2}

\title{Observable primordial vector modes}

\author{Antony Lewis}
 \email{antony@cosmologist.info}
 \affiliation{CITA, 60 St. George St, Toronto M5S 3H8, ON, Canada.}

\begin{abstract}

Primordial vector modes describe vortical fluid perturbations in the
early universe. 
A regular solution exists with constant non-zero
radiation vorticities on super-horizon scales. 
Baryons are
tightly coupled to the photons, and the baryon velocity only decays by an order unity
factor by recombination, leading to an observable CMB anisotropy
signature via the Doppler effect. There is also a large B-mode CMB
polarization signal, with significant power on scales larger than
$l\sim 2000$. This B-mode signature is distinct from that
expected from tensor modes or gravitational lensing, and makes a
primordial vector to scalar mode power ratio $\sim 10^{-6}$ 
 detectable. Future observations aimed at detecting large
 scale $B$-modes from gravitational waves will also be sensitive to
 regular  vector modes at around this level. 
\end{abstract}
\maketitle

\vskip .2in

Observations of the cosmic microwave background (CMB) show that the
primordial perturbation was almost certainly dominated by adiabatic
scalar (density) modes. However it is well known that there are
several possible scalar isocurvature modes~\cite{Bucher99} that
could be present at some level. In the presence of a primordial
magnetic field, there is also an observable vector mode
perturbation~\cite{Subramanian:2003sh,Lewis:2004ef} sourced by the
anisotropic stress of the magnetic field. Other sources such as
topological defects can also source vector modes. Here we concentrate on the
rarely-considered regular
primordial (unsourced) vector modes, which are non-decaying solutions
of the perturbation equations in the presence of free streaming neutrinos~\cite{Rebhan92}.
We show that a  very small primordial regular vector mode amplitude
could be observable. 

In the
absence of an initial large scale radiation vorticity the vector modes
remain in a decaying mode and have essentially no observational
signature. They are therefore not predicted to be present at
any significant level in inflation or other simple models.
 However there is a regular mode with a non-zero initial photon
 vorticity, having equal and
opposite initial photon and neutrino angular momenta such that the
total large scale angular momentum is zero.
This is the
vector analogue of the scalar neutrino isocurvature velocity mode
discussed in Ref.~\cite{Bucher99}, and constitutes a valid possible
component of the general primordial perturbation.
These velocity modes would have to be excited after neutrino decoupling and are
hence difficult to produce and somewhat contrived. But they remain a
logical possibility that can be constrained by observation, and if
observed would be a powerful way to rule out most theoretical models (for
constraints on the scalar mode see e.g. Ref.~\cite{Bucher:2004an} and references therein).
The vector mode can be detected at very small amplitudes and
distinguished from the various scalar modes because of its
distinct non-zero 
$B$-mode (curl-like) CMB polarization signal that is absent with only linear scalar
modes. 

As we show, the vector $B$-mode signature is quite different from that expected from
weak lensing or primordial tensor modes.
On large scales the spectrum is similar to that from tensors, so
observations aimed at detecting the $B$-modes from primordial tensors will also be
sensitive to the large scale part of the vector power spectrum, but
they can easily be distinguished by the vector mode
power on smaller scales. The physical difference between
the spectra is that
tensor modes rapidly decay as soon as they come inside the horizon,
whereas the vortical modes are nearly constant during radiation domination,
decaying only on small scales though damping towards the end of tight coupling.

Even more contrived regular modes exist with non-zero primordial
neutrino octopole (or higher)~\cite{Rebhan92,Rebhan:1994zw}, however these have a much weaker
observational signature and are not considered further here.

\short{}{\subsection{Covariant Equations}}

We consider linear perturbations in a flat FRW universe evolving
according to general relativity with a cosmological constant, neglect any velocity dispersion
of the dark matter and baryon components, and approximate the neutrinos as massless. Perturbations
can be described covariantly in terms of a 3+1 decomposition with
respect to some choice of observer velocity $u_a$ (we use natural
units, and the signature where $u_a u^a=1$), following Refs.~\cite{Ellis83,Gebbie99,Challinor:1998xk}.
Projected spatial derivatives orthogonal to $u_a$ can be used to
quantify perturbations to scalar quantities, for example the pressure
perturbation can be described in terms of $\uD_a p$ where the
spatial derivative is
\ba
\uD_a \equiv \grad_a - u_a u^b\grad_b.
\ea

Conservation of total stress-energy $\grad^a T_{ab}=0$ implies an
evolution equation for the total heat flux $q_a$
\ba
\dot{q}_a + \frac{4}{3} \Theta q_a + (\rho+p)A_a - \uD_a p + \uD^b \pi_{ab}= 0.
\ea
where $\rho$ is the energy density, $\dot{q}_a \equiv u^b\grad_b q_a$,
$\Theta \equiv \grad^a u_a$ is three times the Hubble expansion, 
$A_a \equiv u_b\grad^b u_a$ is the acceleration, and $\pi_{ab} \equiv
T_{\la ab\ra}$ is the total anisotropic stress. Angle brackets around
indices denote the projected symmetric trace-tree part (orthogonal to
$u_a$).

We define the vorticity vector $\Omega_a \equiv  \curl u_a$
where for a general tensor
\ba
\curl X_{a_1\dots a_l} \equiv \eta_{bcd(a_1} u^b \uD^c X^{d}{}_{a_2\dots a_l)}
\ea 
and round brackets denote symmetrization. It has the evolution equation
\ba
\dot{\Omega}_a + \frac{2}{3} \Theta \Omega_a =  \curl A_a
\ea
and is transverse $\uD^a \Omega_a = 0$.
Remaining quantities we shall need are the `electric' $E_{ab}$ and
`magnetic' $H_{ab}$ parts of the Weyl tensor $C_{abcd}$
\ba
E_{ab} \equiv C_{acbd} u^c u^d \quad\quad 
H_{ab} \equiv  \frac{1}{2} \eta_{acdf} C_{be}{}^{cd}u^e u^f
\ea
(which are frame invariant) and the shear $\sigma_{ab} \equiv \uD_{\la a} u_{b \ra}$. The Einstein
equation and the Bianchi identity give the constraint equations
\begin{eqnarray}
\uD^a \sigma_{ab} - \frac{1}{2}\curl \Omega_b - {\frac{2}{3}}\uD_b \Theta - \kappa q_b
= 0 \nonumber\\
\uD^a E_{ab} - \kappa \left({\frac{\Theta}{3}} q_b + {\frac{1}{3}}
\uD_b \rho + {\frac{1}{2}} \uD^a \pi_{ab}\right) =0 \nonumber\\
\uD^a H_{ab} - {\frac{1}{2}}\kappa [ (\rho + p)\Omega_b + \curl q_b ]=0
\nonumber \\
H_{ab} - \curl \sigma_{ab} + \frac{1}{2}\uD_{\langle a} \Omega_{b \rangle}=0,
\label{constraints}
\end{eqnarray}
and the evolution equations
\begin{eqnarray}
\dot{\sigma}_{ab} + \frac{2}{3} \Theta \sigma_{ab} & = &  - E_{ab}
- {\frac{1}{2}}\kappa \pi_{ab} \label{eq:tp3} \nonumber\\
\dot{E}_{ab}+ \Theta E_{ab} & = & \curl H_{ab} + {\frac{\kappa}{2}}
\left[  \dot{\pi}_{ab}  -(\rho + p)\sigma_{ab}
+ {\frac{\Theta}{3}}\pi_{ab}\right] \label{eq:tp4} \nonumber\\
\dot{H}_{ab} +\Theta H_{ab}& = & - \curl E_{ab} -
{\frac{\kappa}{2}}\curl \pi_{ab}.
\label{props}
\end{eqnarray}
We use natural units where $c=1$ and define $\kappa \equiv 8\pi G$.

A vector like $A_a$ may be split into a scalar part $A_a^{(0)}$ and a
vector part $A_a^{(1)}$ where $A_a = A_a^{(0)} + A_a^{(1)}$,
$A_a^{(0)} = \uD_a A$ for some first order scalar $A$ and the vector
part is solenoidal $D^a A_a^{(1)} =0$. This extends to a tensor where
the vector part is given by $\sigma_{ab}^{(1)} = \uD_{\la a}
\Sigma_{b\ra}$ for some first order solenoidal vector $\Sigma_{b}$.

\short{We choose the frame $u_a$ to be
hypersurface orthogonal so that $\curl u_a = 0$ and $(\bar{\uD}_a
X)^{(1)}=0$, 
}{
We can choose $u_a$ to simplify the analysis. 
At linear order one can
always write $u_a = u_a^\perp +
v_a$, where $u_a^\perp$ is hypersurface orthogonal and $v_a$ is first order, so $\curl u_a =
\curl v_a$. For a zero order scalar quantity $X$ it follows that $\uD_a X =
\uD_a^{\perp} X - v_a \dot{X}$. 
For vector modes $(\uD_a^{\perp} X)^{(1)} =0$, and 
it is convenient to choose the frame $u_a$ to be
hypersurface orthogonal so that $\curl u_a = 0$ and hence $(\bar{\uD}_a
X)^{(1)}=0$, 
}
where the bar denotes evaluation in the zero vorticity frame.
From its propagation equation, vanishing of the vorticity also implies that $\bar{A}_a^{(1)}=0$,
so the zero vorticity frame coincides with the synchronous
gauge. The CDM velocity is also zero in this frame modulo a mode which decays as
$1/S$ where $S$ is the scale factor. 


It is convenient to expand the vector components in terms of transverse
eigenfunctions of the zero order Laplacian, $Q^\pm_a$ where $S^2 \uD^2 Q_a^\pm
= k^2 Q_a^\pm$ and $\pm$ denotes the parity. A rank-$\ell$ tensor may
be expanded in terms of rank-$\ell$ eigenfunctions $Q^\pm_{A_l}$ defined by
\ba
Q_{A_l}^\pm \equiv \left(\frac{S}{k}\right)^{l-1} \uD_{\la a_1} \dots \uD_{a_{l-1}} Q_{a_l \ra}^\pm
\ea
which satisfy
\ba
\uD^{a_l} Q^{\pm}_{A_{l-1}a_l} &=&
\frac{k}{S} \frac{(l^2-1)}{l(2l-1)} Q_{A_{l-1}}^\pm
\\
\curl Q_{A_l}^{\pm} &=& \frac{1}{l} \frac{k}{S} Q^{\mp}_{A_l}.
\ea
Harmonic coefficients are defined by
\ba
\sigma_{ab}^{(1)} &=& \sum \frac{k}{S} \sigma Q^\pm_{ab}
\quad\quad
H_{ab}^{(1)} = \sum \frac{k^2}{S^2} H  Q^\pm_{ab}
\nonumber\\
q_a^{(1)} &=& \sum q  Q_a^\pm
\quad\quad
\Omega_a = \sum \frac{k}{S} \Omega Q^\pm_a
\nonumber\\
\pi_{ab}^{(1)} &=& \sum \Pi Q_{ab}^\pm
\ea
where the $k$ and $\pm$ dependence of the harmonic coefficients is
suppressed and $q_i=(\rho_i+p_i) v_i$ for each fluid component, where
$v_i$ is the velocity, and the total heat flux is given by
$q=\sum_i q_i$. The sum is over $k$ and the $\pm$ parities.
We write the baryon velocity simply as $v$.

The equations for the harmonic coefficients in the zero vorticity
frame reduce to
\ba
k(\bsigma' + 2\H \bsigma) = - \kappa S^2\Pi \nonumber\\
H = \half\bsigma \quad\quad\quad 2\kappa S^2 \bq = k^2\bsigma
\ea
where the dash denotes a derivative with respect to conformal time $\eta$,
and $\H = S\Theta/3$ is the comoving Hubble parameter.
The combination $v+\sigma$ (the Newtonian gauge velocity) is frame invariant, as are
$\bsigma = \sigma + \Omega$ and $\bv = v - \Omega$. By choosing to
consider the
zero vorticity frame we have simply expedited the derivation of the
above frame invariant equations. Other papers use
the Newtonian gauge~\cite{Hu:1997hp}, in which $\bsigma$ 
is the vorticity.

The evolution equation for the shear has the solution
\ba
\bsigma = \frac{-1}{S^2} \int d\eta \frac{\kappa S^4 \Pi}{k}.
\ea
In the absence of anisotropic stress it therefore decays as
$1/S^2$. However after neutrino decoupling the neutrinos will supply
an anisotropic source, and solution of this equation requires a
consistent solution for the neutrino evolution.

\short{The
$\ell$-component of the fractional temperature anisotropy $\Delta_l$ is given by
\ba
\Delta_l(\eta_0) = \int^{\eta_0} \!\!\!&d\eta& e^{-\tau}\biggl[ Sn_e \sigma_T \bv
\Psil(\chi)\nonumber \\
&&+ \left(Sn_e\sigma_T \frac{\zeta}{4} +  k \bsigma \right) \frac{d\Psil(\chi)}{d\chi}\biggr]
\ea
Here $\zeta\equiv 3I_2/4 - 9E_2/2$ is a source from the photon
anisotropic stress and $E$-mode polarization, and $\rho_\gamma I_2 = \Pi_\gamma$.
}{

The baryon velocity is coupled to the photon velocity via Thomson
scattering
\ba
\bv' + \H \bv = -\frac{\rho_\gamma}{\rho_b}  S n_e \sigma_T \left(
  \frac{4}{3}v - I_1\right)
\ea
where $I_1 = 4v_\gamma/3$, $n_e$ is the electron number
density and $\sigma_T$ is the Thomson scattering cross-section.

The photon multipole equations~\cite{Challinor:1998xk} for vectors become
\begin{multline}
\bI_l' + k \frac{l}{2l+1}\left[ \frac{(l+2)}{(l+1)} \bI_{l+1} -
  \bI_{l-1}\right] =\\
-S n_e\sigma_T\left( \bI_l - \frac{4}{3}\delta_{l1}\bv -
\frac{2}{15}\zeta \delta_{l2}\right) + \frac{8}{15} k \bsigma \delta_{l2} 
\end{multline}
where $\zeta\equiv 3I_2/4 - 9E_2/2$ is a source from the photon
anisotropic stress and $E$-mode polarization, and
$\rho_\gamma I_2 = \Pi_\gamma$.  In
general $I_l$ is an angular moment of the fractional photon density
distribution, four times the fractional temperature anisotropy.
The neutrino multipole
equations are analogous but without the Thomson scattering terms.
The solution is
\begin{multline}
I_l(\eta_0) = 4\int^{\eta_0} \!\!\!d\eta e^{-\tau}\biggl[ Sn_e \sigma_T \bv
\Psil(\chi) \\
+ \left(Sn_e\sigma_T \frac{\zeta}{4} +  k \bsigma \right) \frac{d\Psil(\chi)}{d\chi}\biggr]
\end{multline}
where $\Psil(x) \equiv l j_l(x)/x$, $j_l(x)$ is a spherical Bessel
function, $\chi \equiv k(\eta_0 -\eta)$ and $\tau$ is the optical depth.
}
In the approximation that the visibility $Sn_e\sigma_T e^{-\tau}$ is a delta function at last
scattering $\eta=\eta^*$ this becomes
\ba
\frac{I_l(\eta_0)}{4} \approx \left[(v+\sigma)\Psil  + \frac{\zeta}{4} \frac{d\Psil}{d\chi}
\right]_{\eta*} \!\!+  2 \int_{\eta*}^{\eta_0} d\eta H' \Psil.
\ea
The anisotropy therefore comes predominantly from the Newtonian
gauge baryon velocity at last scattering, plus an integrated
Sachs-Wolfe (ISW) term from the
evolution of the magnetic Weyl tensor $H=\bsigma/2$ along the line of sight.

The vector polarization multipole equations~\cite{Challinor:2000as} 
\short{have solutions}
{
become
\ba
E^\pm_l{}' &+& \frac{(l+3)(l+2)l(l-1)}{(l+1)^3(2l+1)}  k
E_{l+1}^\pm
 - \frac{l}{2l+1} k E_{l-1}^\pm  \nonumber\\
&-& \frac{2}{l(l+1)} k B^\mp_l = -Sn_e\sigma_T(E_l^\pm - \frac{2}{15} \zeta^\pm\delta_{l2})
\nonumber
\\
B^\pm_l{}' &+& \frac{(l+3)(l+2)l(l-1)}{(l+1)^3(2l+1)} k
B_{l+1}^\pm \nonumber
- \frac{l}{2l+1} k B_{l-1}^\pm \\
&+& \frac{2}{l(l+1)} k E^\mp_l = 0
\ea
where $E_l$ and $B_l$ describe moments of the E (gradient-like) and B
(curl-like) polarization. These equations have solutions
}
\ba
E_l(\eta_0) &=&
 \frac{l-1}{l+1} \int^{\eta_0} d\eta S n_e \sigma_T
e^{-\tau}\left[ \frac{d\Psil(\chi)}{d\chi} + \frac{2\Psil(\chi)}{\chi}\right] \zeta
\nonumber\\
B_l(\eta_0) &=& -\frac{l-1}{l+1} \int^{\eta_0} d\eta S n_e \sigma_T
e^{-\tau} \Psil(\chi) \zeta. 
\ea
\short{}{Signs of $E_l$ and $B_l$ here follow the conventions of CMBFAST~\cite{Seljak:1996is} and CAMB~\cite{Lewis:1999bs}.}

\short{}{\subsection{Solutions}}

\begin{figure}
\begin{center}
\psfig{figure=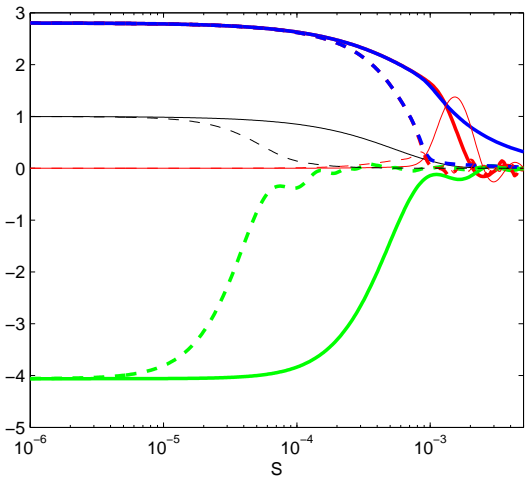,angle=0,width = 7cm}
\caption{
Evolution of the vector perturbations with wavenumber $k=0.02
  \Mpc^{-1}$ (solid lines) and $k=0.2
  \Mpc^{-1}$ (dashed lines).
Thick lines are the
  velocities of the baryons (top), photons (until decoupling same as
  the baryon velocity) and neutrinos
  (bottom). Thin lines are $\bsigma$ (with $\bsigma_0=1$) and the photon
  anisotropic stress. The
  baryon velocity evolves independently of wavenumber on large scales,
  but is damped on small scales.
\label{ev}}
\end{center}
\end{figure}

At early times the baryons and photons are tightly coupled, the
opacity $\opacity \equiv S \text{n}_e \sigma_{\text{T}}$ is
large. This means $v_\gamma \approx  v$, and we
can do an expansion in $\tau_c$ that is valid for
$\epsilon\equiv \max(k\tau_c, \H\tau_c) \ll 1$. To lowest order
\ba
\bv' &=& -\frac{R \H \bv}{1+R} +{\cal O}(\tau_c)
\ea
where $R\equiv 3\rho_b/4\rho_\gamma$. 
The solution is
\ba
\bv \approx \frac{\bv_0}{1+R} 
\ea
where $\bv_0$ is the initial value. Hence if $\bv_0 \ne 0$, by
decoupling $\bv$ has only decayed an order unity factor
depending on the matter and radiation density at the time. 
On smaller scales where $k\tau_c ={\cal O}(1)$ before decoupling the
perturbations are damped by photon diffusion, giving a characteristic
fall off in perturbation power on small scales.


\begin{figure}
\begin{center}
\psfrag{l}[][][1.7]{$\ell$}
\psfrag{Cl}[][][1.7]{$\ell(\ell+1)C_\ell/(2\pi\mu K^2)$}
\psfig{figure=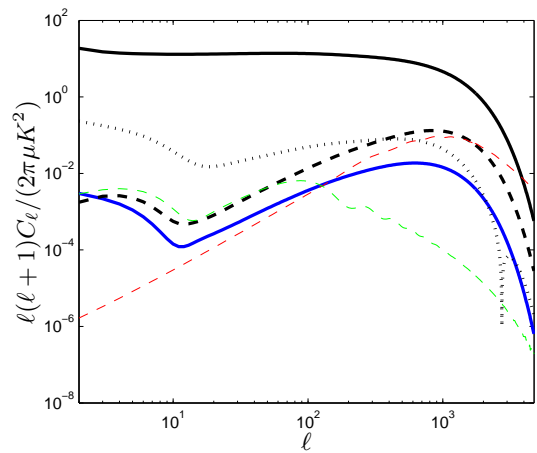,angle=0,width = 7cm}
\caption{
Typical CMB temperature (top solid), polarization EE (bottom solid),
BB (dashed thick) and cross-correlation TE (dotted; absolute value) power
spectra for regular vector modes assuming a primordial vector to scalar power ratio $\sim
10^{-3}$ and scale invariant vector mode spectrum $P_\bsigma$. The other dashed lines
show the $B$-mode spectrum from weak
lensing (peaking at $\ell\sim 1000$), and primordial tensors with initial power
ratio $\sim 10^{-1}$ (peaking at $\ell \sim 100$). 
\label{Cls}}
\end{center}
\end{figure}

We now perform a general series expansion in conformal time for the
above equations in the early radiation dominated era to identify the regular primordial modes. We define $\omega\equiv
\Omega_m\H_0/\sqrt{\Omega_R}$, where $\Omega_R =
\Omega_\gamma+\Omega_\nu$, and $\H_0$ and $\Omega_i$ are the Hubble
parameter and densities (in units of the critical density) today.
\short{}{
The Friedmann equation gives
\ba
S = \frac{\Omega_m \H_0^2}{\omega^2}\left( \omega\eta +
\frac{1}{4}\omega^2\eta^2 +\clo(\eta^5) \right).
\ea
}
Defining the ratios $R_\nu\equiv\Omega_\nu/\Omega_R$, $R_\gamma
\equiv\Omega_\gamma/\Omega_R$, $R_b\equiv\Omega_b/\Omega_m$,
and keeping lowest order terms the regular solution (with zero initial
anisotropies for $l>2$) is
\ba
\bsigma &=& \bsigma_0\left(1 - \frac{15}{2}
  \frac{\omega\eta}{4R_\nu+15}\right) \\
\bv_\gamma &=& \bsigma_0  \frac{4R_\nu+5}{R_\gamma} \left( \frac{1}{4}  -
\frac{3R_b}{16R_\gamma}\omega\eta\right) 
\\
\bv_\nu &=& -\frac{\bsigma_0}{4} \frac{4R_\nu+5}{R_\nu} \\
I_2 &\equiv& \frac{\Pi_\nu}{\rho_\nu} =  - \frac{2}{3}
\frac{k\eta}{R_\nu} \bsigma_0 
\ea
where we have neglected small contributions from the scattering-suppressed
photon anisotropic stress. This regular mode is the vector analogue of the neutrino velocity
isocurvature mode discussed in Ref.~\cite{Bucher99}.
The shear $\bsigma$ is initially constant on super-horizon scales,
supported by the growing anisotropic stress of the neutrinos. On sub-horizon
scales in radiation domination it decays as the neutrino anisotropic stress starts to
oscillate rather than grow.  

The photon and neutrino vorticities are
constant on super-horizon scales during radiation
domination. This is consistent with angular momentum conservation
because of the energy redshift. The photon vorticity is tightly
coupled to the baryons, so both are initially nearly constant, with
some decay 
due to drag from the baryons through 
matter radiation equality. On super-horizon scales there is only an order
unity decay, so a significant large scale photon quadrupole
will be present at low redshift to source a significant additional large
scale polarization signal from reionization.  
The evolution is illustrated in Fig.~\ref{ev}.

On large scales the early ISW contribution is about 20\% as $\bsigma$ decays
as the matter becomes more dominant. On scales sub-horizon at
recombination there is no ISW contribution as
$\bsigma$ has already decayed. We neglect the effect of magnetic field
generation by the photon-baryon vorticity~\cite{Rebhan92}.

\short{}{\subsection{Observations}}

We now compute the observable CMB anisotropy signal. 
We define the dimensionless first order transverse vector $\sigma_a$ such that
$\sigma_{ab}^{(1)} = D_{\la a} \sigma_{b\ra}$, and quantify the
primordial vector modes by their power spectrum $P_{\bsigma}$ defined
so that
\ba
\la |\bsigma_a|^2 \ra = \int d \ln k \,\,  P_\bsigma.
\ea
The corresponding expressions for the CMB temperature and polarization power spectra are
derived in~\cite{Lewis:2004ef}.

To account for the small scale damping effect accurately, as well as a
detailed treatment of recombination and reionization, 
we compute sample CMB power spectra numerically by a straightforward
modification of CAMB\footnote{\url{http://camb.info/}}. The CMB power
spectra ($C_l$)
depend on $ P_\bsigma$. For a scale invariant spectrum,
the temperature $C_l$ has a broad peak around $\ell \sim 50$, as shown in
Fig.~\ref{Cls}. The polarization power spectra peak at around $\ell \sim
500$, with the $B$-mode dominating in accordance with
Ref.~\cite{Hu:1997hp}.  

The large scale reionization signal is rather similar to that
expected from tensor modes, and thus experiments aimed at detecting
this tensor signal will also be sensitive to vector modes. Incomplete sky
coverage only decreases the sensitivity by an order unity factor due
to $E$-$B$ mode mixing~\cite{Lewis01,Lewis:2003an} even on the largest
scales. From
Fig.~\ref{Cls} we see that the large scale $B$-modes are more
sensitive to vector power by a factor of about 100, thus sensitive
observations of tensor modes will also be good probes of regular vector
modes. To distinguish the two one just needs to measure
the spectrum at $\ell \agt 100$ where the tensor power falls but the
vector power continues to grow.

The dominant confusion on small scales is likely to be from
weak lensing of the scalar modes, which peaks on similar
scales. There are about $10^6$ observable modes, so one can ideally expect to
detect a vector contribution $\sim 1/1000$ of the power of the
lensing signal. Since they are of comparable power for a scale invariant
primordial power spectrum ratio $P_\chi/P_\bsigma$ of $\sim 10^{-3}$
($P_\chi$ is the power in the comoving curvature perturbation), this implies that
vector modes with only $10^{-6}$ of the scalar power may be
detectable irrespective of the tensor mode amplitude. Since the lensing signal is non-Gaussian, and in the
absence of vector modes is partially subtractable~\cite{Hirata:2003ka,Kesden:2003cc,Knox:2002pe}, the in-principle
limit is probably much lower, though this depends on the
spectrum of the vector modes. The ultimate limit may be around the
level where there should be a sourced vector mode signal from second order effects~\cite{Mollerach:2003nq}.

Primordial magnetic fields source a $B$-mode spectrum similar to
that from primordial vector
modes~\cite{Subramanian:2003sh}.
However the perturbations are expected to
be highly non-Gaussian for magnetic fields and hence easily distinguishable from
primordial vector modes if they are approximately Gaussian, at least
until the lensing confusion limit. Magnetic fields also provide a
constant source which partly
compensate the damping, so there is more magnetic field vector mode
power on very small scales.
The detailed signature of magnetic fields in the CMB is discussed
in Ref~\cite{Lewis:2004ef}, including the additional large scale
signature from tensor modes. 

Topological defects can also source similar
$B$-mode spectra~\cite{Seljak:1997ii}, though again the spectrum is
expected to be non-Gaussian, and (at least for strings) there is more
power on very small scales due
to the continuous sourcing of the vector modes.

\short{}{\subsection*{Conclusion}}

We have shown that regular primordial vector modes have a strong observational
signature, allowing the possibility that tiny primordial amplitudes can be constrained from
future high-sensitivity CMB polarization $B$-mode observations. Any
signature of vector modes would be powerful evidence against simple
inflationary models. The
Planck\footnote{\url{http://astro.estec.esa.nl/Planck}} satellite
should be able to detect the
$B$-mode signature from primordial vector modes at the $10^{-3}$
level, and distinguish them from tensor modes by the
presence of small scale power.  A full Bayesian joint analysis of all
the CMB power spectra should be straightforward using MCMC techniques,
and may give better constraints that suggested here. Separating
a vector mode signal at the $10^{-6}$ level from that generated by lensing of scalar modes
would be a serious challenge for the future. 

\short{}{\subsection*{Acknowledgements}}

I thank Anthony Challinor for discussion of similar work and valuable
advice, Jochen Weller and Sarah Bridle for useful comments, and Marco Peloso for
stimulating discussions.


\end{document}